# Cell differentiation: what have we learned in 50 years?

Stuart A. Newman[*]

[*]*Department of Cell Biology and Anatomy, New York Medical College, Valhalla, New York, USA*

ABSTRACT

I revisit two theories of cell differentiation in multicellular organisms published a half-century ago, Stuart Kauffman's global gene regulatory dynamics (GGRD) model and Roy Britten's and Eric Davidson's modular gene regulatory network (MGRN) model, in light of newer knowledge of mechanisms of gene regulation in the metazoans (animals). The two models continue to inform hypotheses and computational studies of differentiation of lineage-adjacent cell types. However, their shared notion (based on bacterial regulatory systems) of gene switches and networks built from them, have constrained progress in understanding the dynamics and evolution of differentiation. Recent work has described unique write-read-rewrite chromatin-based expression encoding in eukaryotes, as well metazoan-specific processes of gene activation and silencing in condensed-phase, enhancer-recruiting regulatory hubs, employing disordered proteins, including transcription factors, with context-dependent identities. These findings suggest an evolutionary scenario in which the origination of differentiation in animals, rather than depending exclusively on adaptive natural selection, emerged as a consequence of a type of multicellularity in which the novel metazoan gene regulatory apparatus was readily mobilized to amplify and exaggerate inherent cell functions of unicellular ancestors. The plausibility of this hypothesis is illustrated by the evolution of the developmental role of Grainyhead-like in the formation of epithelium.

## 1. Introduction

In 1969 two highly influential papers were published on the mechanism of cell-type diversification, or cell differentiation, in multicellular organisms such as animals and plants. One of them, published in this journal, Stuart Kauffman's "Metabolic stability and epigenesis in randomly constructed genetic nets" (Kauffman, 1969), was a harbinger of modern systems biology in that it addressed the question with a global dynamical model with distributed causality. The other paper, which appeared in *Science*, was Roy Britten's and Eric Davidson's "Gene regulation for higher cells: a theory" (Britten and Davidson, 1969). This paper was less theoretically ambitious, since its formal structure was based on the modular and hierarchical cybernetics prevalent in the 1950s and 60s. Both the Kauffman "global regulatory genome dynamics" (GGRD) model and Britten-Davidson "modular gene regulatory network" (MGRN) model took their inspiration from the striking advances that had been made over the previous decade in bacterial gene regulation (reviewed in Loison and Morange (2017)). But while the GGRD model aimed to reconceptualize developmental biology, the MGRN model was more concerned with adapting experimental findings in prokaryotic systems to accounting for new findings in multicellular eukaryotes.

A common assumption of the two models, new at the time, but which has been carried forward in contemporary understandings of differentiation, was the inference from cloning experiments in amphibians by John Gurdon and others that terminally differentiated cells retained all the genes necessary for full-term development (reviewed in (Blau, 2014)). All subsequent models of differentiation have been based (with exceptions for some developmental lineages such as the immune system) on differential gene activity rather than theoretically conceivable alternatives like progressive gene loss or gene innovation.

Both models postulated the existence of molecular switches whereby genes of the "regulatory genome" (a more recent coinage, Davidson (2006), but a concept implicit in each framework) control the activity of other genes via the transcription factors (TFs) or (presciently in the case of the MGRN model), the regulatory RNAs they specify. The control is exerted via cis-acting promoters associated with the regulated genes. In the MGRN class of models the TFs at the bottom of the developmental hierarchy in a particular cell type then switch on "batteries" (linked sets) of terminal genes that mediate the cells' characteristic functions (Britten and Davidson, 1969). (How the appropriate spatial partitioning is accomplished was left vague.) In the GGRD class of models, combinations of TFs corresponding to attractors of a genome-wide dynamical system are the signatures of distinct cell types, with the expression of each type's function-mediating genes following automatically from their presence. This model also did not address the question of spatiotemporal allocation of cell fates.

Although features of both frameworks have survived in current experimentally based models of developmental lineage generation (see below), each of these perspectives has been disconfirmed in important ways. The major failure of both derives from their shared conception of the genetic switch. As noted above, this was based on analogies with bacterial gene regulatory processes now recognized to be inapplicable to eukaryotic, and particularly to metazoan cells. The newer understanding of gene control in these more complex organisms and the difficulties it presents for both the GGRD and MGRN frameworks are the main subject of this article. But the two earlier perspectives have other fatal flaws.

The assumption of GGRD-type models that the full complement of cell types of an organism corresponds to the mathematically determined set of attractors of its regulatory genome was implausible from the outset. These regulatory networks would have appeared (in the case of the animals) with the first metazoans, organisms resembling present-day sponges and placozoans, which have less than a dozen cell types. These cell types, mainly epithelial-like (attached directly to one another) and mesenchyme-like (embedded in an extracellular matrix), and stem-cell-like (which give rise to the others), would have needed to have been simultaneously attractors of the regulatory genome of an ancestral animal and functionally compatible with each other in the context of a full organism. As these cell types were complemented in later-appearing forms with additional ones, specifying, for example, muscle cells, neurons, and gland cells, the regulatory genomes of the later-diverging forms would have been required to exhibit attractors corresponding to both the old and the new cell types, as well as their frequently conserved developmental lineages.

Results from comparative developmental biology and phylogenomics has shown that similar sets of TFs often do specify corresponding cell types in increasingly more complex animals. But conservation of the compositional identity of attractors and the hierarchical relationships among them as new variables are added are not generic properties of any class of dynamical systems. While the global attractor idea continued to be affirmed by advocates of the GGRD model for many years after it was proposed ("it is almost an inevitable hypothesis that the distinct cell types of an organism correspond to the distinct attractors of the network" (Andrecut and Kauffman, 2006)), it is no longer part of any empirically informed research agenda.

The MGRN framework, for its part, was initially accompanied by no mathematical model other than a formal logical structure that was a concatenation of bacterial operon-like modules. The presumption was that each cell type was programmed independently and hierarchically. This view was maintained by Eric Davidson through this life. The embryos of complex organisms were held to be "genomic computers" (Istrail et al., 2007). Specific cell types were the outcomes of developmental programs consisting of "directed, oriented networks in which information flows in only one direction" (Peter and Davidson (2015), p. 43). In phylogenetic terms, conserved regulatory states can be successively redeployed as modules for new cell types without the system-level constraints on their generation implied in the evolution of GGRDs (Royo et al., 2011). This computational picture, while consistent with features of some developmental systems, is contradicted by nonhierarchical, nonmodular, multilevel, multifactorial mechanisms found in others (Berkseth et al., 2013; Pombero et al., 2018; Zhang et al., 2016).

Despite the conceptual problems with both the GGRD and MGRN frameworks, current developmental models make use of insights from each. The representation of the regulatory genome by networks of Boolean functions (e.g., logical switches with discrete ON/OFF outcomes) of Kauffman's original and later formulations of the GGRD model (Huang, 2001; Kauffman, 1969) was ultimately adopted by MGRN investigators for analysis of specific modules. This has inevitably made their approach less hierarchical and informationally unidirectional, and therefore more realistic (Cui et al., 2017; Davidson, 2011; Peter and Davidson, 2011).

The more mathematically sophisticated GGRD perspective has long employed (in addition to Boolean networks) representations of genetic networks and switches by systems of ordinary differential equations (ODEs) (Glass and Kauffman, 1973; Keller, 1995). But this more natural-seeming format is analytically and computationally unwieldy in high dimensions. Once the goal of this research program to generate the full panoply of an organism's cell types from its genome was relinquished, however, impressive progress could be made on specific developmental lineages such as the red and white blood cell-forming system (Chang et al., 2006; Mojtahedi et al., 2016). The dynamical + modular compromise between the two 1969 theoretical perspectives (termed here the "consensus model") has become so standard that it is used by default in developmental simulations of systems as varied as insect sensory organs, nematode vulval lineages, and floral morphogenesis (Corson et al., 2017; Corson and Siggia, 2017; Alvarez-Buylla et al., 2008).

What is common to all differentiation models that employ Boolean network or ODE representations of genetic circuitry, however, is a concept of switches based on mechanical, electrical, or chemical processes. This is understandable considering the era in which these models were first advanced, and they appear to reflect the reality of the bacterial gene regulatory systems that inspired them. Since then, however, features of eukaryotic, and specifically metazoan, chromatin organization and gene regulation that distinguish gene regulation in animals from bacterial promoter-based mechanisms, have shown Boolean or ODE dynamics to be inapplicable models for cell differentiation other than in a heuristic sense. Specifically, the classical assumptions that the networks have fixed topology (connections among components), that genes are switched on and off by stoichiometric processes (guaranteeing conservation of mass in a chemical reaction, or corresponding consistency of quantitative effects in a discrete system) that obey mass action (i.e., having predictable rates based on amounts of agents present), are all violated by metazoan mechanisms of developmental gene regulation that have come to light in the past few years.

Two main factors undermine the classical switching network representations. The first is the fact that most developmentally critical TFs (at least 90% of them in animal systems) contain intrinsically disordered protein regions (IDRs), which introduce uncharacterized conditionalities into their binding of DNA sequences and their partnering with cofactors (Liu et al., 2006)). The second is the realization that these structurally variable regulatory molecules exert their control in activating and inhibiting phase-separating and -transforming liquid-state droplets in the cell nucleus, into which enhancer sequences are recruited (up to thousands, also of variable specificity) from distant sites in the genome. This makes GRNs a more nebulous concept than suggested by the switchboard-like circuit diagrams that have become standard in developmental biology (Nicholson, 2019; Niklas et al., 2015).

Since the generation and transformations of cell types during development employ switching events in a formal sense, Boolean networks and ODE systems can serve as "toy models" that capture some of their phenomenological aspects. While this accounts for the success of some of the recent analyses cited above the classical representations are so removed from the physical reality of the processes in question that they fail to capture important biological features. Other mathematical representations, potentially derived from the physics of glasses and other amorphous condensed materials, may ultimately prove more suitable (Li et al., 2017; Tüű-Szabo et al., 2019).

One aspect of cell differentiation which completely eludes both the GGRD and MGRN models is the question of the evolution of cell function, which has attracted increasing attention in the decades since these models were first advanced (Arnellos and Moreno, 2012; Moreno and Mossio, 2015). As noted above, the existence of a dynamical system whose attractors are biologically functional, mutually consistent, and conserved when additional components are added over evolution (the presupposition of the GGRD framework) is entirely implausible. For different reasons, the assumption of stepwise origination (presumably by gradual natural selection) of modular "kernels" (cell type determining GRNs (Davidson and Erwin, 2006)) that are independently deployable in organisms of increasing complexity, also becomes unpersuasive in light of newer knowledge. In contrast to these evolutionary scenarios, recent studies raise the possibility that a global regulatory system, very different from the multi-attractor dynamics of the GGRD model, was capable, in the context of metazoan multicellularity, of appropriating, exaggerating and spatially partitioning single-cell functions of ancestral organisms.

## 2. Metazoan cell differentiation: beyond basal eukaryotic gene regulation

*2.1 Eukaryotic- and holozoan-specific transcriptional mechanisms*

Of the three domains of life, eukaryotes and archaeans have their DNA organized by chromatin-associated proteins (in eukaryotes mainly histones) that are capable of being covalently modified, e.g., by acetylation or methylation, in a way that affects DNA accessibility (Prohaska et al., 2010). In this they are distinguished from bacteria, whose simpler gene control mechanisms served as the prototype for both the GGRD and MGRN frameworks. With chromatin protein modification processes in place, transcriptional programs can be remembered, thus becoming independent of the conditions that first brought them about. Eukaryotes are distinguished from archaeans, however, in having enzymes that can erase and remodel the chemical marks recorded on histones. This provides these organisms, uniquely among cell-based life forms, with "write-read-rewrite" genomic machinery capable of recording and processing past biochemical states (Prohaska et al., 2010).

Nuclei of eukaryotic cells also contain Mediator (mediator of RNA polymerase II transcription; (Verger et al., 2019)), a large multi-subunit protein adaptor complex. Mediator consists of a core of 15 different protein subunits which are rich in IDRs, permitting the complex to potentially interact with thousands of different eukaryotic TFs, which themselves contain IDRs. Mediator transduces signals from upstream cis-acting activating sequences to the transcriptional machinery, assembled at promoters proximal to the target genes.

While the write-read-rewrite machinery and the Mediator transcription-regulating adaptor are conserved throughout the eukaryotes, the evolution of animals involved additional steps. Some of these set Holozoa (a more inclusive clade that contained the unicellular ancestors of the metazoans as well as some extant unicellular and transiently colonial organisms) apart from the rest of the eukaryotes. A multidomain protein complex unique to holozoans, p300/CBP, initiates transcription by acetylating histones and thereby relaxing the structure of nucleosomes at promoters of all or most expressed genes. This facilitates the recruitment of RNA polymerase II and TFs to those sites (Chan and La Thangue, 2001). Extant nonmetazoan holozoans such as *Capsaspora owczarzaki* contain homologs of this complex, and their promoter nucleosomes exhibit p300/CBP-specific histone acetylation marks (Grau-Bové et al., 2017). However, Fungi, which is a sister clade of Holozoans within the broader Opisthokonta, lacks p300/CBP. The closest functional counterpart in yeast is an unrelated protein (Dahlin et al., 2015).

Mediator, p300/CBP and TFs congregate in topologically associating domains (TADs) within the nuclei of holozoans (Furlong and Levine, 2018; Galupa and Heard, 2017; Plys and Kingston, 2018). These are liquid-state protein droplets – "biomolecular condensates" (Alberti et al., 2019; Shin et al., 2019) – that form by phase-separation from the surrounding nuclear sap. In these TADs, p300/CBP serves as a scaffold in which a multitude of TFs and other coregulators such as nuclear receptors (NRs) mediate tissue responses to developmental and physiological cues (Dyson and Wright, 2016). A given gene is expressed when p300/CBP binds to TFs and NRs targeting the gene's promoter, accompanied by TF-dependent phase separation which incorporates Mediator into gene-activating droplets (Boija et al., 2018).

A last important holozoan gene regulatory functionality is the "silencing" of gene expression carried by the histone methylating enzymes SUV39H and the Polycomb Group II proteins (Jih et al., 2017). Though acting on the write-read-rewrite mechanism common to all eukaryotes, silencing by methylation is a synapomorphy of Opisthokonta, having arisen in a common ancestor of holozoans and fungi (Kingston and Tamkun, 2014; Steffen and Ringrose, 2014).

*2.2 Metazoan-specific gene regulatory mechanisms*

While eukaryotic multicellular organisms outside the opisthokonts (most prominently the vascular plants), generate a variety of specialized cell types, among the holozoans only metazoans have this capacity. Cell differentiation in animals, which is the most prolific in all known life forms, is enabled by gene regulatory processes which, while based on the eukaryotic-holozoan platform, have unique properties.

The genes of all nonmetazoan organisms are regulated by upstream cis-acting promoters, but metazoans (with the apparent exception of Placozoa (Sebé-Pedrós et al. (2016)) also employ *enhancers*, promoter-like sequences which can be located upstream or downstream of their target genes, in introns, or even on different chromosomes. Enhancers (based on the definition used in the widely referenced ENCODE Project (ENCODE Project Consortium, 2012), are not found in extant unicellular holozoans and are apparently unique to Metazoa and its immediate antecedents and (Sebé-Pedrós et al., 2016). In contrast to promoters, enhancers are transcribed, the RNAs they specify being integral to their function (Mao et al., 2019). They are also much more numerous than promoters (as many as 50,000 being present in mammalian genomes, for example (Heinz et al., 2015)), which makes them highly suited to mediating the fine-tuned and often high levels of expression of the characteristic genes of terminally differentiated cells (Lenhard et al., 2012; Zabidi and Stark, 2016).

Cell type specification in animals involves several hundred clusters of enhancers, termed super-enhancers (SEs) (Hnisz et al., 2017; (Hnisz et al., 2013, Lovén et al., 2013, Parker et al., 2013, Kundaje et al., 2015, Whyte et al., 2013). The SEs mediate cell differentiation in association with lineage-determining transcription factors (LDTFs) (Heinz and Glass, 2012; Link et al., 2015) in a different manner from the way in which enhancers regulate ubiquitously expressed "housekeeping" genes (Arenas-Mena, 2017; Zabidi et al., 2015). Up to a thousand or more enhancers contained in linearly distant chromatin loops can be recruited to a given cell-type-regulated gene along with p300/CBP and Mediator by such LDTFs as MyoD (for muscle) (Blum and Dynlacht, 2013) and Runx2 (for bone) (Vimalraj et al., 2015) in TAD condensates within the interphase nucleus (Furlong and Levine, 2018; Galupa and Heard, 2017; Plys and Kingston, 2018).

In coordination with the assembly of cell type-specific TADs in response to developmental signals (Arenas-Mena, 2017), genes whose activity would be disruptive to or inconsistent with the function of the cell type are suppressed, ensuring that mixed-identity cell types are not produced (Sunadome et al., 2014). This is accomplished by a novel class of silencers in Metazoa, the PcGI class of proteins (Grossniklaus and Paro, 2014; Steffen and Ringrose, 2014). Unlike the histone methylases found throughout the opisthokonts, these operate antagonistically to LDTFs at the level of biomolecular condensates by a TAD-type phase-separation process (Tatavosian et al., 2019). The incorporation of the suppressed genes into stably inactive heterochromatin and the demarcation of active from inactive condensates is mediated by the mechanochemical activities of the chromatin architectural proteins CTCF and cohesin (Chan et al., 2018; Zheng and Xie, 2019).

This brief overview of the unique metazoan-specific gene regulatory apparatus ("mechanism" seems too rigid a description (Nicholson, 2019)) shows it to be based on fluid-state physical and mechanochemical effects occurring at condensed hubs that are sites of topologically plastic interactions among molecules with conditional identities. This arrangement eludes anything but the most superficial parallels to the operon-based switches that inspired both the GGRD and MGRN models. Further, in contrast to evolutionary enigmas represented by these half-century old frameworks, newer knowledge of animal gene regulation suggests plausible scenarios for the origin of functional differentiation of cell types.

## 3. Metazoan regulatory hubs naturally amplify ancestral cell functions

The metazoan gene regulatory apparatus is particularly suited to amplifying the expression of genes that already perform specific cellular functions. If this was a role it coevolved with, much of the puzzle of the evolution of cell types becomes soluble. According to Arenas-Mena (2017) enhancers evolved in the direct ancestors of metazoans from an interaction of two distinct promoter architectures. Unicellular holozoans have both inducible promoters that respond to external cues and promoters that regulate constitutive genes, i.e., those which are transcribed continually. Transcription factors in nonmetazoan holozoans have the constitutive-type promoter architecture, but Arenas-Mena hypothesizes that some of these TF promoters became responsive to developmental signals by acquiring inducible-type architectures, after which a few small changes converted them into developmental enhancers.

Two kinds of TFs are involved in the amplified gene expression associated with metazoan cell differentiation. The first are "pioneer" TFs (e.g., Oct4, FoxA, Sox2, GATA4) (Iwafuchi-Doi and Zaret, 2016) which open "closed" chromatin for transcription. Most of these were already present (from the evidence of extant species) in premetazoan holozoans. The others are LDTFs (lineage-defining TFs) mentioned above. These act on pioneer TF-capacitated chromatin to initiate expression of suites of genes that ultimately result in one or more terminally differentiated cell types (Obier and Bonifer, 2016). Most LTDFs, e.g., MyoD for skeletal muscle, Nkx3-1for cardiac muscle, neurogenin for neurons, Sox9 for cartilage, (reviewed in Newman et al. (2009), or the gene families that specify them, are also found in nonmetazoan holozoans (Sebé-Pedrós et al., 2011).

The scenario proposed by Arenas-Mena envisions the nonmetazoan prototypes of LDTFs becoming developmentally inducible (Arenas-Mena, 2017). It is reasonable to suggest (though good evidence is lacking) that the ancestral functions of each of those TFs was the regulation of sets of constitutive genes that related to specific single-cell functions. These would have included motility, extracellular matrix production, detoxification, light responsivity, oxygen capture, and lipid storage (but not, for example, photosynthesis), with cell type counterparts in animals of myoblasts, chondrocytes and osteocytes, hepatocytes, retinal rods and cones, erythrocytes, and adipocytes. Thus, although additional evolutionary steps were needed to mobilize inducible TFs to developmental roles, the resulting cell type identities may have been "ready-made" in ancestral cells.

In order for inducible TFs to become LTDFs they would have had to address shared sets of cis-regulatory modules (CRMs), i.e., DNA sequences with binding sites for the signal-induced TFs, or genes whose expression is linked to other genes through preexisting signaling networks (Sonawane et al., 2017). Such coordinated expression of genes, if they mediate a particular function in a limited set of cells in the developing embryo, would have constituted a cell type. In metazoans, cell type-specific regulation depends on proximal promoter-enhancer interactions across distant sites in chromatin. Remarkably, the long-range chromatin associations of functionally related genes seen in animal cells (Cao and Cheng, 2015; Laarman et al., 2019; Singh et al., 2018; Stodola et al., 2018) are also found in fungi (Diament and Tuller, 2017;

Tanizawa et al., 2010). This raises the possibility that the cells that gave rise to the animals were "pre-loaded" with linked sets of genes that served relatively independent physiological activities. This would have enabled the novel metazoan gene regulatory apparatus to readily generate specialized cell types based on accentuation of ancestral functions (reviewed in Newman (2019)).

There is little evidence at present to support this speculation, but one example is tantalizing. Members of the Grainyhead-like (Grhl) family of transcription factors, in an unusual synergy, serve as both pioneer and lineage-determining TFs for epithelium (Jacobs et al., 2018). The epithelial cell is in fact the "ur-cell type" of the metazoans, present in all phyla, including the morphologically simplest (so-called "basal") sponges and placozoans, providing a barrier function seen in all animal species. Grainyhead evolved even before the divergence of holozoans and fungi. In the mold *Neurospora*, for example, the TF GRHL is involved in remodeling the extracellular material between spores, and therefore affects their dispersal (Pare et al., 2012). Significantly, members of the Grhl family of metazoan TFs have a DNA-binding specificity similar to the fungal one, and a related function in modulating cell associations. Grainyhead-like positively regulates E-cadherin and the junctional protein claudin4 (Riesgo et al., 2014; Werth et al., 2010). Simultaneously, it negatively regulates the expression of other genes whose products, metalloproteinases, degrade cadherins, and thereby inhibits cell detachment. The downregulation of gene expression is based on another highly unusual effect of Grhl: depending on its context it can directly inhibit the histone acetylase domain of p300, as well as more typically (for pioneer TFs and LTDFs) activating it (Pifer et al., 2016).

A last exceptional property of Grhl relating to its possible role at the origin of differentiation is its titration of levels of free β-catenin. This protein has a dual role in (i) preserving the integrity of metazoan embryos as an essential component of cadherin-based adhesion and (ii) promoting cell type-specific gene expression. Bound β-catenin remains sequestered in an inner membrane complex with the cytoplasmic domain of classical cadherins, where it consolidates and strengthens cell adhesion. Free β-catenin enters the cell nucleus where it typically associates (in p300/CBP-containing condensates) with TFs of the TCF/Lef family, converting them from repressors into activators of transcription of a wide variety of cell type-specific target genes (Archbold et al., 2012; Najdi et al., 2011). Grainyhead-like's role in upregulating cadherins and (by suppressing metalloproteinases) protecting them from degradation allows it to modulate free β-catenin levels, fine-tuning the balance between tissue cohesion and differentiation, essential to the control of multicellular development.

Grainyhead-like is also a central factor in organotypic cell differentiation later in embryogenesis, consistent with its role in establishing epithelia (e.g., in kidney development Boivin and Schmidt-Ott (2018)), and beyond it. Regarding the latter, a combined assay of functional TF-gene interactions and physical TF-DNA interactions determined that Grhl was bound to thousands of nucleosome-free regions associated with the genes regulated during Drosophila eye development (Potier et al., 2014). It is therefore plausible that Grhl TFs, with their dual pioneer and LDTF roles at the juncture of multicellularity and regulation of specific gene expression, might have been the prototype for the recruitment of ancestral coordinated cell functions to differentiated cell types. While other TFs might be found with similarly concerted roles, it seems unlikely that any individual one would exhibit them all.

**4. Conclusion: cell differentiation as an emergent property of metazoan multicellularity**

As discussed above, some important features of both the global regulatory genome dynamics (GGRD) and modular gene regulatory network (MGRN) models of the mid-20$^{th}$ century have survived as part of a consensus framework for the explanation of metazoan (and higher plant) cell differentiation. In particular, modern heuristic models are both dynamical and reciprocally causal (using Boolean functions or ODEs as per the GGRD concept) and locally modular (as per the MGRN concept). But what we have learned since that time has shown that the mechanisms for switching among cell states and types does not conform to the mechanical or chemical presuppositions of the earlier models, since it utilizes components with variable identities (e.g., TFs with conditional cis-acting binding motifs) organized into topologically plastic networks. While (as we have seen) the generation of an animal's cell types is indeed a global process (as anticipated by the GGRD model) the mode of storage of alternative states is nothing like the attractor sets of a dynamical system hypothesized in the original formulation of this idea.

The most disqualifying aspect of the earlier models, however, is their failure to contain plausible scenarios for how cell differentiation may have evolved. Specifically, what systemic properties of their regulatory genomes provided metazoans with the capability of generating a dozen or so cell types in early branching clades, and then up to several hundred cell types as evolution proceeded? As discussed above, the appearance, postulated by the GGRD model, of a multi-attractor system with functional utility of each separate attractor state and physiological coherence among them, which would then conserve these states as new regulatory genes, controlling new functions, were added over time, was always improbable. It is no surprise, then, that metazoan cell types have turned out not to be attractors of any classically defined unitary dynamical system.

The original MGRN model, for its part, advanced no hypothesis for how modular GRNs arose, and then accumulated, other than tacitly assuming the trial-and-error gradualism of the evolutionary theory of a half-century ago, the Modern Synthesis. Some recent scenarios for the origination and evolution of cell types remain firmly within the MGRN-Modern Synthesis framework ((Arendt et al., 2016; Wagner et al., 2019). These employ logical circuits resembling the combinatorial ones of the Britten-Davidson model, termed "cell type-specific core regulatory complexes" (CoRCs). These comprise sets of transcription factors and their cooperative interactions specifying particular cell types (referred to as "independent gene expression modules" (Arendt et al., 2016)). The combinatorial nature of CoRCs enables diversification (presumably by standard adaptive selection) into sister cell types (e.g., rod and cone photoreceptors of the vertebrate retina). These are embellished and refined by "apomeres," novel functionalities particular to given cell subtypes, involving new proteins or protein-protein interactions (e.g., rod and cone phototransduction systems).

Like the "consensus" mathematical/computational models mentioned in the Introduction, the CoRC-apomere model is a reasonable representation of transitions between adjacent states in developmental lineages, insofar as logical switches can serve as a heuristic for the metazoan differential gene expression amplification apparatus, described in the previous section. However, when an account of the evolution of new CoRCs is sought, neglect of the nature of the actual switching and gene expression amplification process of metazoans leaves this model on uncertain grounds. Acknowledging that "from the few cell types of metazoan ancestors to the many hundred cell types of most extant bilaterians…hierarchical trees, representing cell type evolutionary and developmental lineage histories, may be incongruent" (Arendt et al., 2016) (see for example, Nowotschin et al. (2019)), the authors advance the ad hoc solution of evolutionarily diversified "serial sister cell types," which may appear to be the same, but are regulated differently in different tissues and regions of the body.

In contrast, the plasticity of the metazoan gene expression apparatus described here suggests a ready solution to the variability of regulatory modes of functionally defined cells in different tissues. In particular, it easy to see how CoRCs of sister cell types could be context- and condition-dependent. But to acknowledge this would raise the possibility that the cell type-originating CoRCs themselves might have been emergent physiological manifestations of inherent cell functions, mobilized automatically by the novel gene regulatory system that appeared coincidently with metazoan multicellularity. This is consistent with the fact that most LTDFs were present in holozoan ancestors (Sebé-Pedrós et al., 2011), as are (in unicellular opisthokonts) function-related, chromatin coregulatory associations (Diament and Tuller, 2017; Tanizawa et al., 2010). This scenario, however, would represent a significant departure from the conventional evolutionary picture implicit in the CoRC model (Arendt et al., 2016; Wagner et al., 2019).

Regardless of the eventual resolution of these conceptual uncertainties, it is unquestionable that we are at the threshold of a completely new era in the understanding of cell differentiation. The recognition that eukaryotic gene expression mechanisms are entirely different from those of the bacterial systems that inspired both the GGRD and MGRN models, and that metazoan expression mechanisms, though based on the former, are even more different from them, has upended the traditional concept of the gene switch. The MGRN model never contained a hypothesis for the origination of cell types beyond the standard gradualism of the Modern Synthesis. And while the GGRD went beyond that in identifying cell types as examples of the "order for free" available in self-organizing systems (Kauffman, 1995), the type of multi-attractor dynamical system used to model differentiation was not an apt one.

While it may eventually turn out that cell types are the behavioral modes a single physical system, as suggested by the GGRD, it will be a system embodying phase transformations of protein-nucleic acid complexes, the principles of which defy familiar laws. In the words of the mid-20[th] century biologist E.E. Just, written 30 years before the GGRD and MGRN models were proposed, understanding development will require "a physics and chemistry in a new dimension …superimposed upon the now known physics and chemistry" (Just, 1939). We will see what the next decades bring.

Zabidi, M. A., Stark, A., 2016. Regulatory enhancer-core-promoter communication via transcription factors and cofactors. Trends Genet 32, 801-814, doi:10.1016/j.tig.2016.10.003.

Zabidi, M. A., Arnold, C. D., Schernhuber, K., Pagani, M., Rath, M., Frank, O., Stark, A., 2015. Enhancer-core-promoter specificity separates developmental and housekeeping gene regulation. Nature 518, 556-9, doi:10.1038/nature13994.

Zhang, L., Liu, W., Zhao, J., Ma, X., Shen, L., Zhang, Y., Jin, F., Jin, Y., 2016. Mechanical stress regulates osteogenic differentiation and RANKL/OPG ratio in periodontal ligament stem cells by the Wnt/beta-catenin pathway. Biochim Biophys Acta 1860, 2211-9, doi:10.1016/j.bbagen.2016.05.003.

Zheng, H., Xie, W., 2019. The role of 3D genome organization in development and cell differentiation. Nat Rev Mol Cell Biol, doi:10.1038/s41580-019-0132-4.